# Frustrated $J_1$-$J_2$ Diamond Lattice Antiferromagnet $Co_2Ti_3O_8$ with a Vacancy-ordered Spinel Structure Synthesized via a Topochemical Reaction

Rio Kumeda,[1] Yuya Haraguchi,[1,*] Daisuke Nishio-Hamane,[2] Akira Matsuo,[2] Koichi Kindo,[2] and Hiroko Aruga Katori[1]

[1]*Department of Applied Physics and Chemical Engineering, Tokyo University of Agriculture and Technology, Koganei, Tokyo 184-8588, Japan*
[2]*The Institute for Solid State Physics, The University of Tokyo, Kashiwa, Chiba 277-8581, Japan*
[*]Corresponding author: chiyuya3@go.tuat.ac.jp

Metastable $Co_2Ti_3O_8$ was synthesized through a topochemical reaction using $Li_2CoTi_3O_8$ as the precursor, resulting in a vacancy-ordered spinel structure. Crystal structure analysis confirmed that Co ions selectively occupy the A-site, giving rise to a frustrated diamond lattice. Magnetic susceptibility and heat capacity measurements revealed antiferromagnetic order at 4.4 K, which is markedly suppressed compared to the negative Weiss temperature of ∼ −27 K, indicating a high degree of frustration effects. Pulsed high-field magnetization measurements revealed a four-step successive magnetic phase transition, demonstrating that $Co_2Ti_3O_8$ is a promising candidate for a frustrated $J_1$-$J_2$ diamond lattice. Additionally, the $J_2/J_1$ ration estimated from the molecular field approximation suggests the possibility of a spiral ordered ground state. These observations highlight the potential of frustrated magnetism in ordered spinel structures to expand the material search space for quantum magnetism, including magnetic skyrmions.

## I. INTRODUCTION

Magnetic skyrmions, which are vortex-like spin texture in solids, have gained considerable attention in condensed matter physics because of their topological protection and their manipulability by electric currents, making them promising candidates for novel magnetic memory and logic devices [1-7]. To stabilize these spin texture, specific conditions in the crystal structure— most importantly, the breaking of inversion symmetry —are necessary [8,9]. In non-centrosymmetric crystals, the Dzyaloshinskii–Moriya interaction governs the orientation electron spins [10,11], leading to the vortex-like spin arrangement characteristic of magnetic skyrmions. These skyrmions have been observed predominantly in non-centrosymmetric systems, such as chiral-lattice magnetic materials. However, the limited range of available materials continues to restrict progress in this rapidly developing research field.

A promising strategy for stabilizing magnetic skyrmions without relying on crystal symmetry involves exploiting magnetic frustration [12]. Magnetic frustration arises when competing magnetic interactions cannot be satisfied simultaneously, leading to stable yet intricate spin textures. By carefully engineering materials to include these frustrated interactions, it becomes possible to realize robust skyrmion states, thus providing a versatile alternative to conventional symmetry-dependent mechanisms. Notably, frustration-driven skyrmions have been observed in various crystal lattices, such as triangular lattices [13,14], breathing kagome lattices [15], square lattices [16], and diamond lattices [17]. These discoveries underscore the significance of frustration-induced phenomena and bolster expectations for further advances in the study of frustrated magnetic materials.

The A-site in $AB_2X_4$ spinel compounds, which crystallize in the $Fd\bar{3}m$ space group, forms a bipartite diamond lattice. This structure can exhibit magnetic frustration arising from competing nearest-neighbor interaction $J_1$ and next-nearest-neighbor interaction $J_2$. One of the most well-known members among these highly frustrated A-site spinel family is $MnSc_2S_4$, where $Mn^{2+}$ ion with spin $S = 5/2$ form a spiral spin structure [17-20]. Very recently, $MnSc_2S_4$ was also found to host a magnetic field-induced antiferromagnetic skyrmion lattice state [17].

A-site spinels present numerous opportunities for the discovery of exotic and novel properties, both experimentally and theoretically. To date, A-site spinels have demonstrated a wide range of intriguing behaviors in various systems, including the cobaltates $Co_3O_4$ [21], $CoRh_2O_4$ [22], aluminates $MAl_2O_4$ ($M$ = Mn, Co) [23-26], and the thiospinel $MnSc_2S_4$ [17-20].

Among them, $Co_3O_4$, $CoRh_2O_4$, and $MnAl_2O_4$ exhibit relatively low frustration indices, $f = |\theta_W|/T_{AFM} = 3.7$, 1.2, and 3.6, respectively, placing them well within the Néel phase. By contrast, $CoAl_2O_4$ appears more strongly

frustrated ($f = 8 \sim 22$), yet its frustration effect and magnetic ground state remain disputed controversial because of unavoidable cation-mixing of Co and Al. This cation mixing is also highly sample-dependent [27], obscuring efforts to clarify the underlying ground state. Meanwhile, MnSc$_2$S$_4$ with $f = 10$ hosts an exotic spiral spin liquid state and serves as a skyrmion platform [28-31]. These observations underscore the importance of developing new materials to advance understanding of frustrated diamond-lattice systems. However, in recent years, reports of new $AB_2X_4$ spinel compounds have been scarce, indicating a stagnation in materials exploration. Research efforts have therefore extended beyond the conventional spinel framework to highly distorted diamond lattices, including the distorted spinel $A$Rh$_2$O$_4$ ($A$ = Cu, Ni) [22,32,33], and the $\gamma$-LiFeO$_2$-type lanthanoid oxides such as LiYbO$_2$ [34, 35] and NaCeO$_2$ [35], where spiral magnetic order has also been observed.

Under these circumstances, we focus on the diamond lattice formed by $A$-sites in spinel structures that have an ordered arrangement of $B$-sites. In these spinel oxides, the $B$ sublattice adopts a pyrochlore lattice, consisting of corner-sharing tetrahedra. When each tetrahedron in the $B$ sublattice is occupied by different ions in a 3:1 ratio, the resulting network forms a three-dimensional array of corner-sharing triangles known as a hyperkagome lattice. This ordering pattern exhibits chirality, with two degenerate chiral structures, $P4_132$ and $P4_332$. For example, Zn$_2$Mn$_3$O$_8$ [36] and Zn$_2$Ti$_3$O$_8$ [37] are classified as chirality-ordered spinel compounds. Specifically, in Zn$_2$Ti$_3$O$_8$, the Ti ions and vacancies are arranged in an ordered manner to form a Ti hyperkagome lattice, while the divalent Zn ions establish a diamond lattice.

Hypothetically, in Co$_2$Ti$_3$O$_8$ [Fig. 1(b)], an analog of Zn$_2$Ti$_3$O$_8$ [37], Co$^{2+}$ occupies the $A$-site presenting a promising candidate for realizing an $S = 3/2$ diamond lattice. However, using QUANTUM ESPRESSO package [38], our calculations indicate that Co$_2$Ti$_3$O$_8$ thermodynamically decomposes into CoTiO$_3$ and TiO$_2$ as follows:

$$\text{Co}_2\text{Ti}_3\text{O}_8 \rightarrow 2\ \text{CoTiO}_3 + \text{TiO}_2, \quad (1)$$

with the decomposition enthalpy of $\Delta H = -157$ kJ/mol. This result indicates that the synthesis of Co$_2$Ti$_3$O$_8$ via conventional solid-state reactions is highly challenging. Consequently, diamond lattice magnets based on ordered spinel structures, such as Co$_2$Ti$_3$O$_8$, have not been synthesized due to their inherent thermodynamic instability. Therefore, achieving successful synthesis necessitates the development of novel kinetic reaction processes.

In this study, we successfully synthesized a metastable Co$_2$Ti$_3$O$_8$ through a topochemical exchange reaction, using Li$_2$CoTi$_3$O$_8$ [39] as a precursor [Fig. 1(a)]:

$$\text{Li}_2\text{CoTi}_3\text{O}_8 + \text{CoSO}_4 \rightarrow \text{Co}_2\text{Ti}_3\text{O}_8 + \text{Li}_2\text{SO}_4. \quad (2)$$

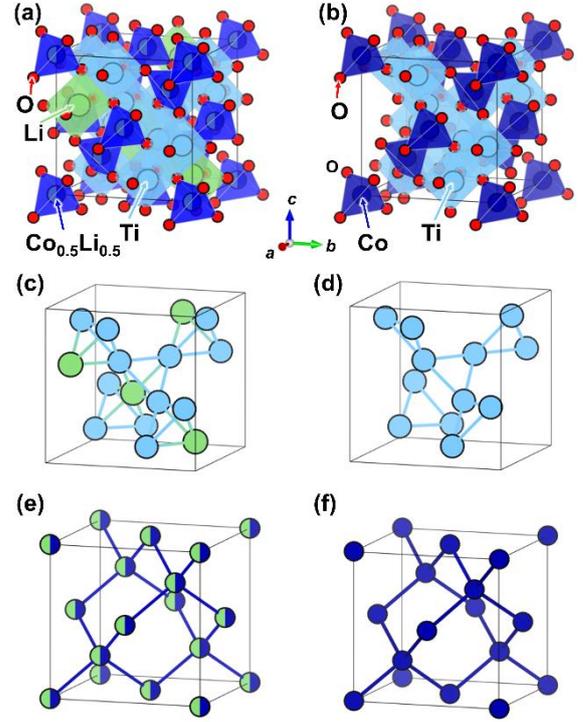

FIG. 1. (a) Crystal structure of the precursor Li$_2$CoTi$_3$O$_8$ and (b) its topochemically transformed product Co$_2$Ti$_3$O$_8$, both adopting space group $P4_132$. Sky-blue and green octahedra represent TiO$_6$ and LiO$_6$ units, respectively; blue and dark-blue tetrahedra denote mixed (Li/Co)O$_4$ and pure CoO$_4$ units. During the Li$_2$CoTi$_3$O$_8$ $\rightarrow$ Co$_2$Ti$_3$O$_8$ transformation, Li ions at octahedral sites are removed, leaving vacancies, while the tetrahedral sites become fully occupied by Co ions. (c, d) Octahedral framework: Li$_2$CoTi$_3$O$_8$ (c) exhibits a pyrochlore-type network in which LiO$_6$ and TiO$_6$ octahedra are ordered in a 1:3 ratio; the Ti$^{4+}$ ions thereby form a continuous hyper-honeycomb lattice. In Co$_2$Ti$_3$O$_8$ (d) the Li sites are vacant, leaving an ordered array of TiO$_6$ octahedra with the same hyperkagome connectivity. (e, f) Tetrahedral framework: mixed Li/Co occupancy in Li$_2$CoTi$_3$O$_8$ (e) is replaced by complete Co filling in Co$_2$Ti$_3$O$_8$ (f), generating a fully occupied Co$^{2+}$ diamond lattice. The crystal structures were visualized using VESTA program [40].

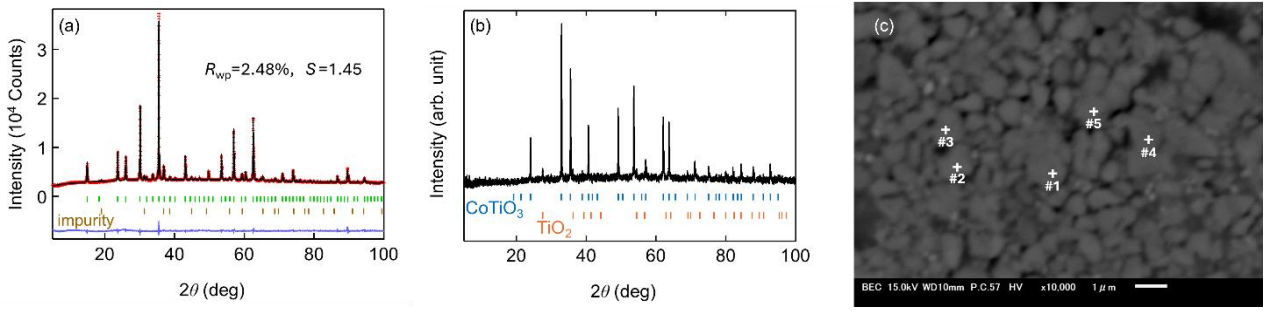

FIG. 2. (a) XRD pattern of a polycrystalline sample of $Co_2Ti_3O_8$. The observed intensities (red circles), calculated intensities (black lines), and their difference (blue lines) are shown. The green and ochre vertical bars indicate the Bragg peak positions of $Co_2Ti_3O_8$ and impurities, respectively. In this refinement, the small amount of impurity in the sample was identified as $Co_3O_4$ with an estimated mass fraction of 6.9 wt%. However, since no magnetic anomaly was observed at the Néel temperature of 40 K for $Co_3O_4$ [42] in the magnetic susceptibility measurement described in the main text, it is estimated that the actual impurity is a spinel-structured nonmagnetic material with a lattice constant close to that of $Co_3O_4$. (b) XRD pattern of the $Co_2Ti_3O_8$ sample annealed at 900°C for 10 h, demonstrating that the $Co_2Ti_3O_8$ sample decomposes into $CoTiO_3$ and $TiO_2$ at such high temperatures. (c) A backscattered electron image of the polycrystalline sample of $Co_2Ti_3O_8$. Numbers indicates EDX measurement spots.

The reaction enthalpy, calculated by QUANTUM ESPRESSO package, is $\Delta H = -66$ kJ/mol, indicating that this designed reaction is highly exothermic and thus proceedable. The precursor $Li_2CoTi_3O_8$ features Li and Co ions each occupying half of the $A$-site [Fig. 1(e)], with the $B$-site ordered by Li and Ti in a 1:3 ratio forming a Ti hyperkagome lattice [Fig. 1(c)]. Crystal structure analysis of the resulting $Co_2Ti_3O_8$ revealed that Co ions selectively occupy the $A$-site [Fig. 1(f)], while the $B$-site is ordered with vacancies and Ti, maintaining the Ti hyperkagome network [Fig. 1(d)], through the topochemical process. The chemical robustness of Ti hyperkagome network is crucial for the formation of $Co_2Ti_3O_8$. Magnetic susceptibility and heat capacity measurements indicated that $Co_2Ti_3O_8$ exhibits an antiferromagnetic order at $T_{AFM} = 4.4$ K, which is relatively suppressed relative to the Weiss temperature $\theta_W \sim -27$ K. This suppression suggests the presence of a considerable frustration effect on a $Co^{2+}$ diamond lattice. Furthermore, pulsed high-field magnetization measurements revealed a four-step successive magnetic phase transition. Our findings highlight $Co_2Ti_3O_8$ as a promising candidate for a frustrated $J_1$-$J_2$ diamond lattice.

## II. EXPERIMENTAL METHODS

The precursor $Li_2CoTi_3O_8$ was obtained using a conventional solid-phase reaction method. $Li_2CO_3$, CoO, and $TiO_2$ were mixed in a molar ratio of 1.05:1:3 and heated in air at 850 °C for 16 h, with a heating and cooling rate of 100 °C/h.

A modified low temperature topochemical synthesis, originally described by Kitani et al. for obtaining $Zn_2Mn_3O_8$ [36], was adapted to synthesize $Co_2Ti_3O_8$ from $Li_2CoTi_3O_8$. In the original procedure, $Li_2ZnMn_3O_8$ (isomorphic to $Li_2CoTi_3O_8$) reacts with $ZnSO_4$ to yield the ordered spinel $Zn_2Mn_3O_8$ [36]. Extending this strategy, two Li ions in $Li_2CoTi_3O_8$ were replaced by one Co ion from $CoSO_4$ via two successive heat treatments, ensuring completion of the cation exchange. First, a 5:1 molar ratio of $CoSO_4$ to $Li_2CoTi_3O_8$ was mixed with 92.32 wt% NaCl (based on the $CoSO_4$ content) as an inert salt, then heated at 230 °C for five days. After the reaction, excess $CoSO_4$, the byproduct $Li_2SO_4$, and NaCl were removed by washing with distilled water. In the second step, the washed sample was again mixed with an equal mass of $CoSO_4$ and 92.32 wt% NaCl, heated at 230 °C for ten days, then washed with distilled water and dried at room temperature, resulting in a polycrystalline product.

The thus-obtained sample was characterized by powder x-ray diffraction (XRD) measurements using a diffractometer with Cu-K$\alpha$ radiation, and chemical analyses were conducted using a scanning electron microscope (SEM; JEOL IT-100) equipped with an attachment for energy-dispersive X-ray spectroscopy (EDX; 15 kV, 0.8 nA, 1 μm beam diameter) at The Institute for Solid State Physics (ISSP), the University of Tokyo. The crystal structure was refined through Rietveld analysis employing the Z-RIETVELD software [41].

The temperature dependence of magnetic susceptibility was measured under several magnetic fields using the Magnetic Property Measurement System (MPMS; Quantum Design) at ISSP, the University of Tokyo. The temperature dependence of the heat capacity was measured using the thermal relaxation method with the Physical Property Measurement System (PPMS; Quantum Design) at ISSP, the University of Tokyo. To estimate the lattice contributions, we measured the heat capacity of the non-magnetic analog Zn$_2$Ti$_3$O$_8$. The isothermal magnetization curves up to $\mu_0 H$ = 50 T at 1.3 K and 4.2 K were measured by the induction method in a multilayer pulsed magnet at the International Mega Gauss Science Laboratory in ISSP, The University of Tokyo.

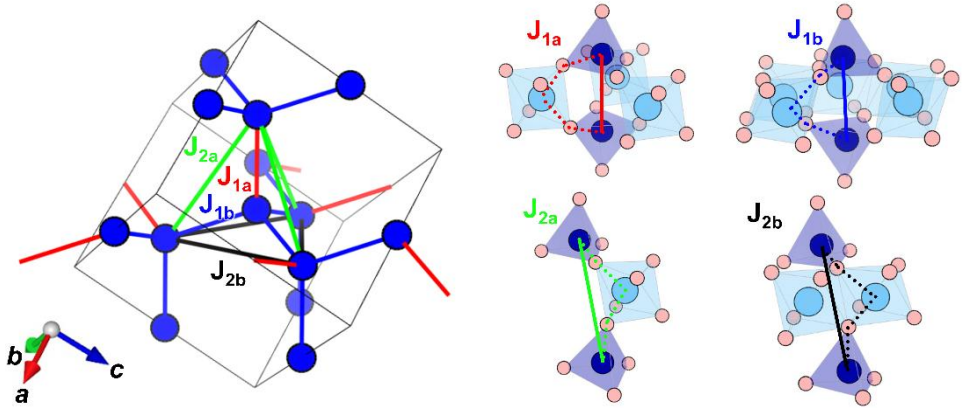

FIG. 3. Left: Diamond lattice of Co$^{2+}$ ions in Co$_2$Ti$_3$O$_8$ highlighting the four crystallographically distinct exchange interactions: $J_{1a}$ (red), $J_{1b}$ (blue), $J_{2a}$ (green), and $J_{2b}$ (black) with the cubic unit cell. Symmetry lowering splits the single nearest-neighbour interaction $J_1$ of an ideal diamond lattice into $J_{1a}$ and $J_{1b}$, and the next-nearest-neighbour interaction $J_2$ into $J_{2a}$ and $J_{2b}$. Right: For each interaction, a representative super-exchange pathway Co–O–Ti–O–Co is illustrated with dashed lines inside TiO$_6$ octahedra (sky-blue) and CoO$_4$ tetrahedra (dark blue). In the actual structure these exchanges occur through multiple equivalent paths: three for $J_{1a}$, five for $J_{1b}$, one for $J_{2a}$, and two for $J_{2b}$. The crystal structures were visualized using VESTA program [40].

Density-functional calculations were carried out with the QUANTUM ESPRESSO package [38]. We employed projector-augmented-wave (PAW) pseudopotentials generated for the PBEsol exchange–correlation functional, applied a Hubbard correction of $U_{\text{eff}}$ = 5 eV on Co 3$d$ states, and used a plane-wave energy cut-off of 60 Ry. Brillouin-zone integrations were performed on a Monkhorst–Pack grid of 11 × 11 × 11 $K$-points, and self-consistency was achieved to better than 10$^{-9}$ Ry in total energy. The density of states was obtained with the tetrahedron method using the same k-mesh, and atomic projections were computed by projecting the Kohn–Sham states onto spherical harmonics inside the PAW augmentation spheres.

### III. RESULTS

The precursor Li$_2$CoTi$_3$O$_8$ sample shows a vibrant light blue color, which changes to a dark bluish green after topochemical reaction. Figure 2 (a) shows the powder XRD patterns of repeatedly reacted CoSO$_4$ +

TABLE I. Refined crystal structure parameters of Co$_2$Ti$_3$O$_8$ (space group: $P4_132$) determined by Rietveld analysis of powder XRD data. The obtained lattice parameter is $a$ = 8.3949(1) Å. 'Occ.' is the site occupancy, and '$B$' is the atomic displacement parameter.

| Atom  | Site | Occ.  | x           | y           | z           | B (Å$^2$)   |
|-------|------|-------|-------------|-------------|-------------|-------------|
| Co(1) | 8$c$ | 1.000 | 0.24691(9)  | = $x$       | = $x$       | 0.81 (2)    |
| Co(2) | 4$a$ | 0.000 | 7/8         | = $x$       | = $x$       | 1.0         |
| Ti    | 12$d$| 1     | 1/8         | 0.86360(8)  | 0.11360(8)  | 0.54 (3)    |
| O(1)  | 8$c$ | 1     | 0.11009(28) | = $x$       | = $x$       | 1.01 (4)    |
| O(2)  | 24$e$| 1     | 0.10999(23) | 0.39200(29) | 0.36036(20) | = $B_{\text{O}(1)}$ |

Li$_2$CoTi$_3$O$_8$ mixture. Excess CoSO$_4$, byproduct Li$_2$SO$_4$, and inert additive NaCl were removed by washing with distilled water. All peaks, except for the impurity peak indexed by the space group $Fd\bar{3}m$, can be indexed using the Zn$_2$Mn$_3$O$_8$ structure [36] (space group $P4_132$, or $P4_332$) with a cubic lattice constant of $a = 8.3949(1)$ Å. This indicates the realization of a 1:3-ordered spinel structure.

Quantitative EDX carried out on five independent crystallites yields an average cation inventory of Co = 2.06(19) and Ti = 2.97(10) atoms per formula unit. Within experimental error these values reproduce the ideal Co : Ti = 2 : 3 ratio required for Co$_2$Ti$_3$O$_8$. Because Li ($Z = 3$) lies below the detection limit of EDX (~0.2 wt %), any residual lithium would necessarily lower the measured Co or Ti counts and thus disturb this stoichiometry; no such deviation is observed. The data therefore place an upper bound of ≤ 0.02 Li atoms per formula unit, indicating that the topochemical de-lithiation is essentially quantitative and that the product is a phase-pure Co$_2$Ti$_3$O$_8$.

In the Rietveld refinement based on the refinement using $P4_132$, the site mixing parameter η, representing the distribution of Co ions between their ideal tetrahedral sites and the nominally vacant octahedral sites, was treated as a free parameter. The preliminary structure of the main phase was set as Co$_{2(1-\eta)}$$^{tet}$[Ti$_3$Co$_{2\eta}$]$^{oct}$O$_8$, and this parameter converged to η = 0, indicating an ideal crystal structure without cation-mixing. The crystal structure parameters obtained for the Co$_2$Ti$_3$O$_8$ phase are summarized in Table I. These structural changes resemble those observed during the topochemical transformation from Li$_2$ZnMn$_3$O$_8$ to Zn$_2$Mn$_3$O$_8$. Although Co and Ti have similar electron numbers, making it difficult to exclude Co–Ti cation mixing definitively based on Rietveld refinement of XRD data, several findings strongly support an ordered arrangement of Co and Ti. First, the vacant site remains unoccupied. Second, the 2:3 ratio of Co to Ti is preserved. Third, to the best of our knowledge, no reports have documented Ti occupying tetrahedral sites in a spinel structure. Finally, both magnetization and heat-capacity measurements exhibit clear evidence of long-range magnetic ordering. Altogether, these observations imply that the Co and Ti sites are essentially fully ordered, though a small degree of cation mixing cannot be conclusively ruled out.

Co$_2$Ti$_3$O$_8$ crystallizes in a chiral cubic lattice that can adopt one of two mirror-related space groups, $P4_132$ or $P4_332$; they are structurally identical except for their handedness. Because laboratory powder X-ray diffraction averages over Friedel pairs, it cannot tell these two enantiomorphs apart, so our Rietveld refinement was carried out in the $P4_132$ setting purely for convenience. This choice does not imply that the material is necessarily "left-handed": the specimen could be a single enantiomer of $P4_132$, the opposite enantiomer $P4_332$, or a racemic mosaic containing domains of both.

Figure 2 (b) shows the XRD pattern of the sample obtained by annealing the resulting Co$_2$Ti$_3$O$_8$ sample at 900°C for 10 h. This pattern confirms the complete decomposition into CoTiO$_3$ and TiO$_2$. This observation demonstrates that Co$_2$Ti$_3$O$_8$ is a metastable phase, kinetically stabilized by topochemical reactions. The refined crystal structure of Co$_2$Ti$_3$O$_8$ is shown in the left part of Fig. 3, and the two types of Co-O distances within the distorted CoO$_4$ tetrahedron are listed in Table II. In contrast to the conventional spinel structure with space group $Fd\bar{3}m$, Co$_2$Ti$_3$O$_8$ crystallizes in the lower symmetrical space group $P4_132$. In this structure, the pyrochlore site of the original compound is replaced by a well-defined 1:3 arrangement of vacancies and titanium ions, causing the titanium sublattice to assume a hyperkagome framework. Because of the symmetry-reduction induced by the ordering of the pyrochlore site, the ideal diamond lattice undergoes a symmetry reduction, causing the single cobalt site typically observed in $Fd\bar{3}m$ spinels into two distinct crystallographic sites. Additionally, one vertex of each CoO$_4$ tetrahedron is elongated, and the TiO$_6$ octahedron also becomes distorted.

In the regular diamond lattice, each Co ion is magnetically coupled to its four nearest-neighbor Co ions by a super-exchange interactions $J_1$, mediated by O and Ti ions via the Co-O-Ti-O-Co superexchange pathways. In Co$_2$Ti$_3$O$_8$ (space group $P4_132$), two distinct types of $J_1$ interactions can be identified based on the differences in bond angles: $J_{1a}$, characterized by the pathway Co–O(2)–Ti–O(2)–Co, and $J_{1b}$, characterized by the pathway Co–O(1)–Ti–O(2)–Co [right part of Fig. 3]. Similarly, the set of twelve next-nearest-neighbor interactions $J_2$ in the regular diamond lattice are also reclassified into $J_{2a}$ and $J_{2b}$ based on their respective pathways. This asymmetry in the lattice structure should affect the spin model.

Figure 4 shows the temperature dependence of the magnetic susceptibility $\chi$ measured at $\mu_0H = 0.1$ T for Co$_2$Ti$_3$O$_8$, where panel (a) now presents the inverse susceptibility $1/\chi$, panel (b) plots $\chi$ together with its derivative d$\chi$/d$T$, and panel (c) contrasts zero-field-cooled (ZFC) and field-cooled (FC) data taken at $\mu_0H = 10$ mT with the FC curve at 0.1 T. No magnetic anomalies attributable to impurities, including the initially suspected Co$_3$O$_4$ suggested by XRD analysis, were observed in the magnetization measurements. The

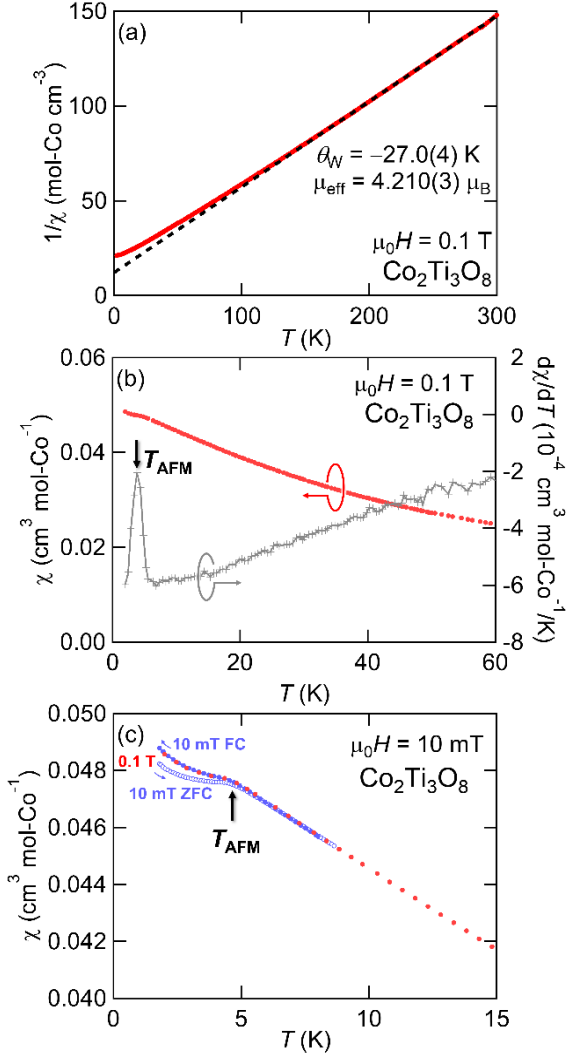

FIG. 4. (a) Temperature dependence of the inverse magnetic susceptibility, $1/\chi$ at $\mu_0 H = 0.1$ T, for $Co_2Ti_3O_8$; the dashed black line represents a Curie–Weiss fit. (b) Magnetic susceptibility chi below 60 K, together with its temperature derivative $d(\chi)/dT$. (c) Zero-field-cooled (ZFC, blue open circles) and field-cooled (FC, blue filled circles) susceptibilities measured in an applied field of $\mu_0 H = 10$ mT; FC data collected at $\mu_0 H = 0.1$ T are shown in red for comparison. The vertical arrow indicates the antiferromagnetic transition temperature, $T_{AFM}$.

Curie–Weiss fitting of the inverse susceptibility yields an effective magnetic moment $\mu_{eff} = 4.210(3)\ \mu_B$ and the negative Curie-Weiss temperature $\theta_w = -27.0(4)$ K for $Co_2Ti_3O_8$. This estimated $\mu_{eff}$-value is consistent with the theoretical one (3.87 $\mu_B$) for a spin quantum number $S = 3/2$, yielding $g \approx 2.17$ assuming $L = 0$. This $g$-value is typical for tetrahedrally coordinated $Co^{2+}$, like $g \approx 2.18$ for $CoRh_2O_4$ [22] and $g \approx 2.26$ for $CoAl_2O_4$ [27].

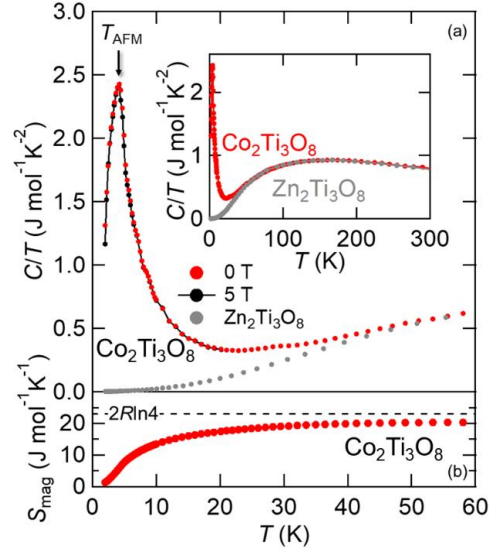

FIG. 5. (a) Temperature dependence of the heat capacity divided by temperature $C/T$ of $Co_2Ti_3O_8$. The red and black circles in the $C/T$ data represent measurements taken at 0 T and 5 T, respectively, and the arrow indicates the magnetic transition temperature $T_{AFM}$. The nonmagnetic analog $Zn_2Ti_3O_8$ is also plotted. Data up to 300 K are shown in the inset. (b) Magnetic entropy $S_{mag}$ obtained by integrating a magnetic contribution of heat capacity in $Co_2Ti_3O_8$.

Combined structural and magnetic analysis confirms that all $Co^{2+}$ ions remain divalent, and the dominant antiferromagnetic interactions among them account for the observed behavior. In contrast, the precursor compound $Li_2CoTi_3O_8$, which also exhibits a negative Weiss temperature (see Supplementary Material [43]), does not exhibit long-range magnetic order. This absence of ordering is attributable to the fact only 50% of the A-sites in the diamond lattice are occupied by $Co^{2+}$, positioning $Li_2CoTi_3O_8$ near the diamond-lattice percolation threshold of ~ 0.43 [44]. The resulting dilute $Co^{2+}$ network prevents the formation of a system-spanning magnetic cluster, thereby suppressing long-range order. Nevertheless, the emergence of a spin-glass-like state at temperatures lower than those examined in the present magnetic susceptibility measurements cannot be ruled out.

As described in Supplemental Materials [43], the change in magnetic moment per chemical formula unit compared to the precursor indicates that the number of Co ions has doubled according to the designed topochemical reaction (2). Additionally, the increase in the absolute value of the Weiss temperature suggests

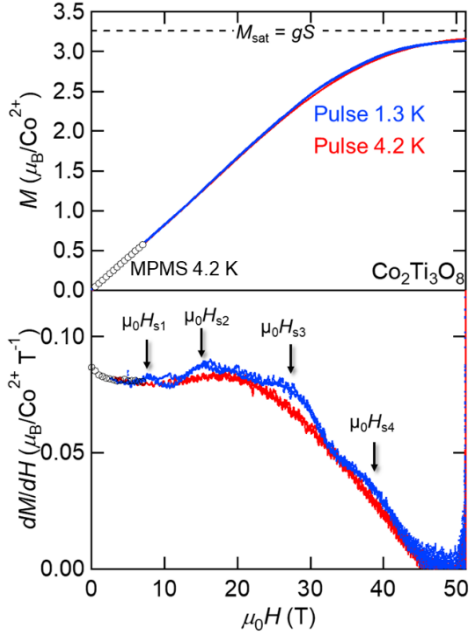

FIG. 6. The isothermal magnetization curve $M$ and their derivatives $dM/dH$ measured in pulsed magnetic field up to 50 T at 4.2 K (red) and 1.3 K (blue) for $Co_2Ti_3O_8$. The magnitudes are calibrated to the data measured under static fields up to 7 T (open circles). The arrows indicate the critical fields.

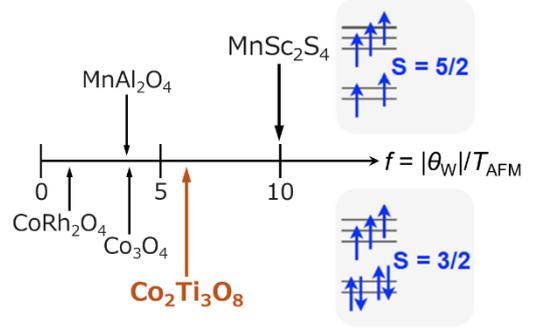

FIG. 7. Comparison of the frustration indices $f = |\theta_W|/T_{AFM}$ in the various diamond lattice antiferromagnets. The $f$-values except for $Co_2Ti_3O_8$ are reproduced from Refs. 19-20, 22-24.

that the increase in the number of magnetic ions through cation exchange effectively increases the pathways contributing to the superexchange interactions between spins.

Furthermore, the $1/\chi$ data deviates from the value expected from the Curie–Weiss law at high temperatures, starting at around 50 K, signaling the growth of antiferromagnetic short-range correlations (see the slight downward curvature in Fig. 4(a)). At lower temperatures, a magnetic anomaly appears at $T_{AFM} \sim 4.4$ K, marked by a pronounced peak in the $d\chi/dT$ curve [Fig. 4(b)], indicating an antiferromagnetic phase transition. The splitting of zero-field-cooled (ZFC) and field-cooled (FC) curves below $T_{AFM}$ under a low magnetic field of $\mu_0H = 10$ mT [Fig. 4(c)], also suggests the occurrence of magnetic ordering. The difference between ZFC and FC magnetization at 2 K is approximately 1.2% of the total magnetization, indicating that this hysteresis does not represent a contribution from the majority of spins.

Figure 5(a) shows the temperature dependence of the heat capacity divided by the temperature $C/T$ for $Co_2Ti_3O_8$. A clear peak is observed at the transition temperature $T_{AFM} = 4.4$ K, as estimated from magnetic susceptibility measurements, suggesting the presence of second order magnetic phase transition with bulk characteristics. Even at an applied magnetic field of 5 T, the peak remains almost unchanged, indicating that the magnetically ordered state is robust against the magnetic field.

Accompanied by the formation of magnetic ordering, the change in the magnetic entropy $S_{mag}$ occurs. To estimate $S_{mag}$, we measured the heat capacity of non-magnetic analog $Zn_2Ti_3O_8$ to determine the lattice contribution of $Co_2Ti_3O_8$. The good agreement between the specific heats of $Co_2Ti_3O_8$ and $Zn_2Ti_3O_8$ above 60 K indicates that $Zn_2Ti_3O_8$ provides the lattice contribution for $Co_2Ti_3O_8$. The magnetic entropy $S_{mag}$ was obtained by integrating the difference between the total specific heat and the lattice specific heat with temperature assuming the $C_{mag}/T$ value equals 0 at 0 K following the third law of thermodynamics. As shown in Fig.5(b), the asymptotic value of $S_{mag}$ at high temperature roughly coincides with the expected total magnetic entropy for $S = 3/2$, $2R\ln(2S+1) = 2R\ln 4 = 23.04$ J mol$^{-1}$ K$^{-1}$ (where $R$ is the gas constant), demonstrating that subtraction of the lattice contribution is valid. The $S_{mag}$-value reaches 6.45 J mol$^{-1}$ K$^{-1}$ at $T_{AFM}$, which is 28% of the total magnetic entropy. This indicates that a large part of the magnetic entropy has been released by the short-range magnetic correlations present above $T_{AFM}$, suggesting that the effect of spin frustration is considerable.

To obtain further insights into magnetic ordering and interactions, magnetization measurements up to 50 T were performed using pulsed magnetic fields, as shown in Fig. 6. At 4.2 K, the magnetization curve shows no significant magnetic anomalies, and no peak structure is observed in the differential magnetization $dM/dH$. In contrast, the $dM/dH$ curve at 1.3 K, which is adequately

lower than $T_{AFM}$, displays peaks at $\mu_0H_{s1}$ = 7.6 T and $\mu_0H_{s2}$ = 15.0 T. Additionally, the differential magnetization exhibits a broad peak at approximately $\mu_0H_{s3}$ = 27.1 T, after which it gradually decreases, tending towards magnetization saturation. A small undulation is observed at $\mu_0H_{s4}$ ~ 39 T, just before the magnetization saturation. These magnetic anomalies are confirmed by differential values derived from raw data obtained in isothermal magnetization measurements using pulsed magnetic fields. Each anomaly is consistent with both increasing and decreasing magnetic field processes. Similar field-induced phase transitions have been reported in a skyrmion host spinel $MnSc_2S_4$ with a diamond lattice [45]. At an applied field of 50 T, the magnetization per $Co^{2+}$ asymptotically approaches the saturation magnetization $M_{sat} = gS = 3.26\ \mu_B$, further corroborating the spin-3/2 state realized in a tetrahedrally coordinated $Co^{2+}$ ion.

## IV. DISCUSSION

As described in the previous section, $Co_2Ti_3O_8$ prepared by topochemical reactions exhibits magnetic ordering. The frustration index $f$, defined as $|\theta_w|/T_{AFM}$, is ~ 6.1, indicating relatively strong frustration even in the absence of cation-mixing. Compared to other antiferromagnets with a diamond lattice structure, as shown in Fig. 7, $Co_2Ti_3O_8$ displays a level of frustration comparable to that observed in $MnSc_2S_4$, a well-known diamond lattice spiral magnet.

We now discuss the chemical trends in frustration across the diamond-lattice spinel magnets, with special emphasis on $Co_2Ti_3O_8$. The decisive variable is whether the non-magnetic $B$ ion offers low-lying empty $d$ orbitals that overlap in energy and symmetry with the anion $p$ band: when such acceptor states are present, anion $p$ electrons acquire an efficient second-order $p$-$d$-$p$ hopping pathway—a perturbative term that selectively boosts the antiferromagnetic next-nearest exchange $J_2$ while leaving the nearest-neighbor term $J_1$ largely unchanged, pushing the system deep into the frustrated regime. This situation is realized in $MnSc_2S_4$ and, most strikingly, in our vacancy-ordered $Co_2Ti_3O_8$, where three $Ti^{4+}$ ions per four $B$ sites supply an abundance of empty $d$ levels. The projected density of states (pDOS) for $Co_2Ti_3O_8$, presented in Fig 8, displays a broad Ti 3$d$ – O 2$p$ hybrid band at the Fermi level, the expected fingerprint of an enhanced $J_2$; an analogous Sc 3$d$ – S 3$p$ hybridization is seen in the $MnSc_2S_4$ pDOS included in the Supplemental Materials [43]. By contrast, $MnAl_2O_4$, whose $B$ ion contributes only deep $p$ orbitals, and the oxides $Co_3O_4$ and $CoRh_2O_4$, whose $B$-site $t_{2g}$ shells are already filled, lack this low-energy channel:

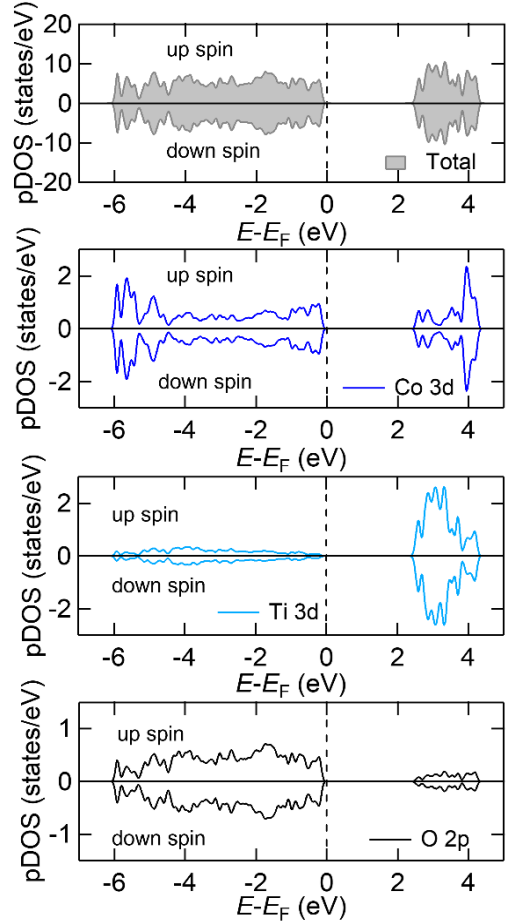

FIG.8. Projected density of states for $Co_2Ti_3O_8$ obtained from DFT + $U$ calculations using projector-augmented-wave (PAW) pseudopotentials with the PBEsol exchange–correlation functional ($U_{eff}$ = 5 eV on Co 3$d$). The total DOS is shown in gray, Co 3$d$ states in blue, Ti 3$d$ states in skyblue, and O 2$p$ states in black.

the virtual-hopping amplitude collapses, $J_2$ shrinks relative to $J_1$, and magnetic order emerges promptly. Complementary pDOS plots for $MnSc_2S_4$, $MnAl_2O_4$, $Co_3O_4$, and $CoRh_2O_4$, also provided in the Supplemental Materials [43], reproduce the expected hierarchy of $p$-$d$ mixing and thereby corroborate this mechanism. A quantitative extraction of the $d$–$p$ energy separation and

TABLE II. Two Co-O distances of distorted $CoO_4$ tetrahedra obtained from the structural analysis. Each oxygen number corresponds to the site number described in TABLE I.

| bond | interatomic distance (Å$^2$) |
|---|---|
| Co–O(1) ×1 | 1.989(3) |
| Co–O(2) ×3 | 1.927(3) |

first-principles exchange constants will be the subject of future work, but the qualitative correspondence already established between electronic structure and magnetic response offers a possible "blueprint" for designing highly frustrated spinel magnets.

Next, we discuss the magnetic properties arising from competing superexchange interactions in $Co_2Ti_3O_8$. Theoretical calculations have been performed on the most stable magnetic structures in the classical $J_1$-$J_2$ model of a frustrated diamond lattice. For small $J_2$, the system exhibits a collinear Néel state. However, when $J_2/J_1$ exceeds 1/8, the ground state becomes highly degenerated, taking the form of conformal helix with a wavevector confined to the helix plane [28]. Here, we inductively estimate the magnetic ground state of $Co_2Ti_3O_8$ by determining the $J_2/J_1$ values for the assumed magnetic structure via the molecular field approximation and then compare those values with the theoretical magnetic phase diagram [28]. Because the $CoO_4$ tetrahedron shows negligible distortion (see TABLE 2), we simplify the calculation by setting $J_{1a} = J_{1b} = J_1$ and $J_{2a} = J_{2b} = J_2$. Within the molecular field approximation, the Néel temperature $T_N$ and Weiss temperature $\theta_w$ are described by:

$$T_N = C(-z_1 J_1 + z_2 J_2) = C(-4J_1 + 12J_2), \quad (3)$$

$$\theta_W = C(z_1 J_1 + z_2 J_2) = C(4J_1 + 12J_2), \quad (4)$$

where $C = 2S(S+1)/3k_B$ and the coordination numbers for the coordination numbers for $J_1$ and $J_2$ are 4 and 12, respectively [33]. Solving these equations using the measured $T_N = 4.4$ K and $\theta_w = -27.0$ K gives $J_2/J_1 = 0.24$. However, this value exceeds 1/8, placing it outside the predicted Néel order region for the $J_1$-$J_2$ diamond-lattice model and indicating that $Co_2Ti_3O_8$ as may line beyond the simple Néel ordering regime.

We next derive the transition temperature for the helical phase using the same general approach. Separate molecular field approximation calculations are conducted for realistic magnetic structures with propagation vectors $(q, 0, 0)$ and $(q, q, 0)$. Although a helical phase characterized by a $(q, q, q)$ wavevector is also considered, the plane perpendicular to k in that case is not symmetric, requiring an additional angle to represent the sublattice phase difference. Consequently, we chose not to perform calculations for that scenario.

Focusing on the $(q, 0, 0)$ direction, propagation vector, four spins linked by $J_1$ rotate by an angle $\varphi$ relative to a central spin, while twelve spins linked by $J_2$ are divided into eight spins that rotate by $2\varphi$ and four spins that remain aligned with the central spin. In the molecular field approximation, the transition temperature $T_q$ is then given by

$$T_q = C(-4J_1 \cos\varphi + 4J_2 + 8J_2 \cos 2\varphi), \quad (5)$$

where $\varphi = 0$ corresponds to the Néel-ordered phase. Substituting $\varphi = 0$ indeed recovers the same expression as in Eq. (3). Minimizing $T_q$ in Eq. (5) with respect to $\varphi$ yields

$$T_q = \frac{C}{J_2}\left(-\frac{J_1^2}{4} - 4J_2^2\right) \quad (6)$$

under the condition $\cos\varphi = J_1/(8J_2)$. Assuming a helical order in $Co_2Ti_3O_8$ with a wavevector along $(q, 0, 0)$, we solve Eqs. (4) and (6) using $T_q = 4.4$ K and $\theta_w = -27.0$ K. These resulting solutions, $J_2/J_1 = 0.15 \pm i0.31$, are complex, implying that there is no real solution in this scenario. Therefore, the magnetic ground state of $Co_2Ti_3O_8$ is not consistent with helical magnetism of the form $(q, 0, 0)$.

For a wavevector in the $(q, q, 0)$ direction, two of the four spins connected by $J_1$ to the central spin rotate by an angle $\varphi$ in different directions, while the remaining two spins remain parallel to the central spin. Among the twelve spins linked by $J_2$, two maintain alignment with the central spin, eight rotate by $\varphi$, and the remaining two rotate by $2\varphi$. Within the molecular field approximation, the transition temperature $T_{qq}$ is then given by:

$$T_{qq} = C(-2J_1 - 2J_1\cos\varphi + 2J_2 + 8J_2\cos\varphi + 2J_2\cos 2\varphi). \quad (7)$$

Substituting $\varphi = 0$ into this expression recovers Eq. (3). Minimizing $T_{qq}$ with respect to phi yields the condition $\cos\varphi = (J_1 - 4J_2)/4J_2$, leading to:

$$T_{qq} = \frac{C}{J_2}\left(\frac{3J_1^2}{4} - 6J_2^2\right). \quad (8)$$

We assume a $(q, q, 0)$ helical order in $Co_2Ti_3O_8$ and solve Eqs. (4) and (8) with $T_q = 4.4$ K and $\theta_w = -27.0$ K, obtaining two real solutions for $J_2/J_1$. Given that magnetization measurements suggest both $J_1$ and $J_2$ are antiferromagnetic ($J_1 < 0$ and $J_2 < 0$), the physically meaningful solution is $J_2/J_1 = 0.519$. This value places $Co_2Ti_3O_8$ in the range $1/2 < J_2/J_1 < 2/3$, consistent with the $(q, q, 0)$ propagation vector region in the theoretical phase diagram [28]. This agreement confirms coherence between our assumptions and the calculated results.

Once the approximations $J_{1a} = J_{1b}$ and $J_{2a} = J_{2b}$ are lifted and higher-order interactions are considered, a more complex magnetic ordering may emerge.

Addressing this possibility requires a clear determination of the magnetic structure, and we are therefore planning neutron diffraction measurements. However, current synthesis methods do not support the large-scale production of samples for neutron experiments, making it challenging to obtain reliable data. Consequently, reevaluating and refining these synthesis protocols remains a critical objective.

Moreover, in the $P4_132$ or $P4_332$ ordered spinel system, the absence of spatial inversion symmetry in the crystal structure could facilitate the stabilization of magnetic skyrmions. It would therefore be valuable to explore whether a skyrmion phase is realized among the high-field magnetic phases. To definitively establish the presence of skyrmions, single crystals are essential, representing a significant experimental challenge that calls for novel crystal growth techniques. Future theoretical studies will likely focus on understanding the stability of skyrmion structures in $P4_132$ or $P4_332$ ordered spinel magnets, where both broken spatial inversion symmetry and frustration interplay.

## V. SUMMARY

We successfully synthesized an $S = 3/2$ frustrated antiferromagnet $Co_2Ti_3O_8$ with a diamond lattice using a topochemical reaction. Crystal structure analysis confirmed that the octahedral coordination sites of the spinel structure, comprising vacancies and $Ti^{4+}$ ions in a 1:3 ratio, crystallize in the $P4_132$ or $P4_332$ space group. Magnetization and heat capacity measurements revealed an antiferromagnetic transition at $T_{AFM} = 4.4$ K, which is worth noting lower than the Weiss temperature $\theta_w$, indicating moderately frustrated magnetism with a frustration index of $f = 6.1$. Furthermore, the isothermal magnetization under pulsed high magnetic fields showed a field-induced successive phase transition, suggesting strong frustration originating from the $J_1$-$J_2$ diamond lattice. Molecular field approximation calculations further suggest the possibility of realizing spiral magnetic order in $Co_2Ti_3O_8$. These findings identify the ordered spinel $A_2B_3X_8$ framework as a promising platform for the study of frustrated diamond-lattice antiferromagnets, particularly due to its magnetic $A$-site with spatial inversion symmetry breaking and spin-frustration. Future work includes systematic material exploration within the $A_2B_3X_8$ family and theoretical studies on $J_1$-$J_2$ type frustration, which may uncover novel phenomena and deepen our understanding of frustrated spin systems.

## ACKNOWLEDGEMENT

This work was supported by JST PRESTO Grant Number JPMJPR23Q8 (Creation of Future Materials by Expanding Materials Exploration Space) and JSPS KAKENHI Grant Numbers. JP23H04616 (Transformative Research Areas (A) "Supra-ceramics"), JP24H01613 (Transformative Research Areas (A) "1000-Tesla Chemical Catastrophe"), JP22K14002 (Young Scientific Research), and JP24K06953 (Scientific Research (C)). Part of this work was carried out by joint research in the Institute for Solid State Physics, the University of Tokyo (Project Numbers 202211-GNBXX-0034, 202212-MCBXG-0011, 202211-HMBXX-0022, 202306-GNBXX-012, 202306-MCBXG-0070, and 202306-HMBXX-0072).